\documentclass[conference]{IEEEtran}
\IEEEoverridecommandlockouts
\usepackage[table]{xcolor}
\usepackage{algorithm}
\usepackage{lipsum}  
\usepackage{cite}
\usepackage{dblfloatfix}
\usepackage{float}
\usepackage{comment}
\usepackage{amsmath,amssymb,amsfonts}
\usepackage{mathtools}
\usepackage{algorithm}
\usepackage{steinmetz}
\usepackage[noend]{algpseudocode}
\usepackage{graphicx}
\graphicspath{{./Figures/}}
\usepackage{textcomp}
\usepackage{siunitx}
\usepackage{url}
\usepackage{array,multirow}
\usepackage[table]{xcolor}
\usepackage{xcolor,colortbl}
\def\BibTeX{{\rm B\kern-.05em{\sc i\kern-.025em b}\kern-.08em
T\kern-.1667em\lower.7ex\hbox{E}\kern-.125emX}}

\begin{document}	
\title{Grid Services by Behind-the-Meter Distributed Energy Resources: NY State Grid Case Study}

\author{\IEEEauthorblockN{Hossein Hooshyar, Rahul Kadavil, Victor Paduani}
\IEEEauthorblockA{Advanced Grid Innovation Laboratory for Energy} 
New York Power Authority, White Plains, NY\\victor.daldeganpaduani@nypa.gov
\and
\IEEEauthorblockN{Aboutaleb Haddadi$^{\dagger}$, AHM Jakaria$^{\ddagger}$, Aminul Huque$^{\ddagger}$}
\IEEEauthorblockA{$^{\dagger}$Transmission Ops. and Planning, $^{\ddagger}$DER Integration}
Electric Power Research Institute, Knoxville, TN\\ahaddadi@epri.com
\and

\IEEEauthorblockN{George Stefopoulos}
\IEEEauthorblockA{\centerline{Boston Government Services, Washington, DC}\\gstefopoulos@bgs-llc.com}

\thanks{This  research  is  supported  by  the  U.S.  Department  of  Energy’s  Office  of Energy  Efficiency  and  Renewable  Energy  (EERE)  under  the  Solar  Energy Technologies Office Award Number DE-EE0009021.}
}

\maketitle

\begin{abstract}

\textbf{This paper presents a case study for utilizing behind-the-meter (BTM) distributed energy resources (DERs) to provide grid services when controlled by a DER Management System (DERMS). The testbed consists of a 5,000 buses transient-stability (TS) real-time (RT) model, two 9,500 buses distribution feeders from local utilities modeled in a distribution system simulator (DSS), and hundreds of DERs. MQTT communication protocol is utilized to interface the models in RT. Two main studies are carried. In the first, it is found that DERs with frequency-watt droop response can help maintaining stability in the future NYS grid in which thermal synchronous generators have been substituted by renewable energy resources. In the second, results demonstrate that BTM DERs can provide similar level of frequency regulation services expected from large utility-scale generation.}

\end{abstract}


\section{Introduction}

\begin{figure}[htb]
    \centering
    \includegraphics[width=0.5\textwidth]{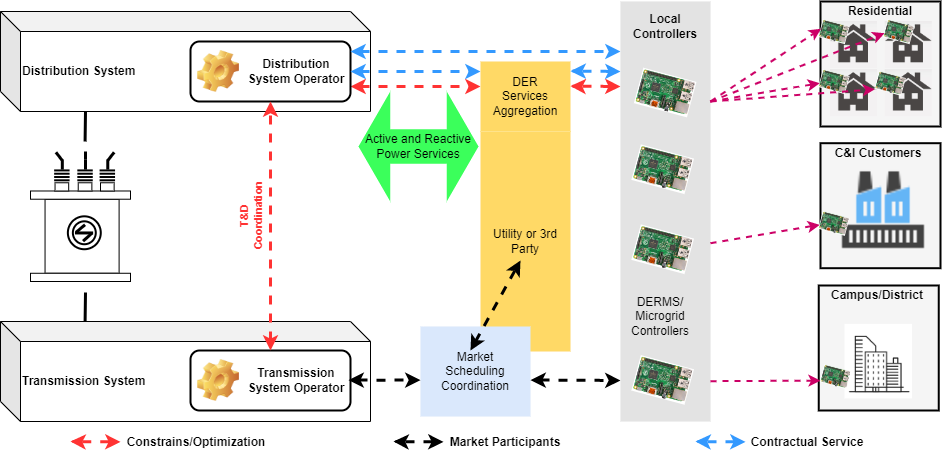}
    \caption{DERMS framework proposed in \cite{aminul2021distributed} as part of the `\textit{Enable BTM DER-provided Grid Services that Maximize Customer Grid Benefits}' (ENGAGE) project.}
    \label{fig:engage_intro}
\end{figure}

To empower distributed energy resources (DERs) to come online, enhance competition, and drive down consumer costs, FERC Order 2222 has been established to enable behind-the-meter (BTM) DERs to provide ancillary services and participate in wholesale energy markets \cite{cano2020ferc}, \cite{EPRI2022FERC}. Achieving this goal entails controlling tens of thousands of devices utilizing various communication protocols. An envisioned framework is to group BTM DERs by local DER management systems (DERMS), which can also be combined to form aggregators \cite{aminul2021grid}. An example of that arrangement is displayed in Fig. \ref{fig:engage_intro}. Moreover, since interconnection standards and requirements such as IEEE 1547-2018 \cite{ieee1547} and California Rule 21 \cite{calrule21} require DERs to be equipped with grid-support functionalities (GSFs), grid operators can leverage complying DERs to assist the grid by actively managing them via DERMS. Therefore, by acting as an interface between operators and small devices, DERMS can help the reliability and economic efficiency of the future grid \cite{aminul2021distributed}. 


Traditionally, transmission and distribution system analysis has been performed using different categories of power system software tools, due to the distinctive nature of power system phenomena studied. However, with increased uptake of DERs equipped with GSFs, the interaction between distribution and transmission system needs to be incorporated in the analysis \cite{matevosyan2021future}. Hence, there has been a growing need for the development of modeling tools capable of co-simulating both transmission and distribution systems including their interactions during transients \cite{jain2021integrated}. Given the long practice of having different time scale tools for transmission and distribution system analysis, it is desired for such a co-simulation platform to be capable of interfacing traditional transmission and distribution tools, while enabling data exchange and time synchronization between the two platforms. For instance, in \cite{palmintier2017design}, researchers have developed a tool for interfacing multiple power system softwares into a single co-simulation environment. Furthermore, real-time (RT) co-simulations are powerful tools for validating the integration of DERs in large grids with hardware-in-the-loop simulation capability, which can provide a realistic representation of how a device would behave in the field. 

A challenge associated with the development of a RT transmission and distribution (T\&D) co-simulation platform is the management of data exchange and time synchronization between the interfaced transmission/distribution tools \cite{poudel2022modeling}. In addition, there is still a lack of available case studies that investigate RT co-simulation models combined with DERMS. Therefore, in this study, we expand the RT T\&D co-simulation framework introduced in \cite{paduani2022real} by integrating the DERMS algorithm from \cite{aminul2021distributed} into the co-simulation loop. The new proposed framework is implemented in a large-scale system for analyzing the capability of BTM DERs in providing grid services such as frequency-watt droop under contingency scenarios and Automatic Generation Control (AGC) regulation signal tracking. The paper presents the following contributions to the literature.


\begin{itemize}
    \item Introduce a real-time (RT) T\&D co-simulation testbed that combines a transient-stability (TS) system, a distribution system simulator (DSS), and a DERMS responsible for controlling BTM DERs with GSFs to provide grid services.
    \item Investigate the impact of having BTM DERs equipped with frequency-watt droop response during a contingency scenario applied to the bulk power system (BPS).
    \item Compare the performance between utility-scale DERs and BTM DERs when tracking an AGC regulation signal from PJM.
\end{itemize}
 
\section{Methodology}

\begin{figure}[htb]
    \centering
    \includegraphics[width=0.5\textwidth]{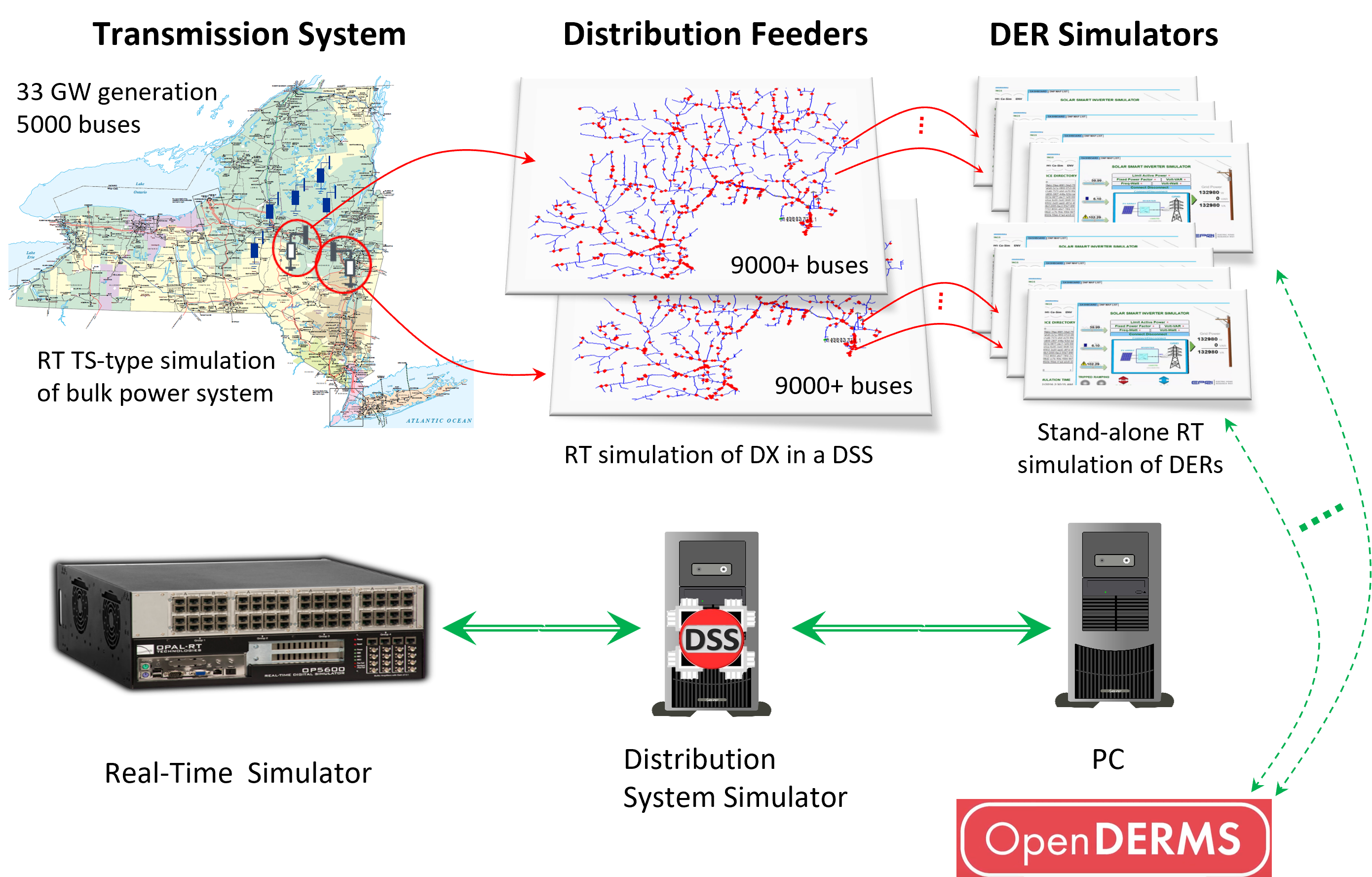}
    \caption{Proposed T\&D co-simulation framework.}
    \label{fig:cosim_struct}
\end{figure}

The co-simulation structure is presented in Fig. \ref{fig:cosim_struct}. The system consists of (i) a section of the Eastern Interconnection Transmission System (EITS), preserving only the NYS grid, with 33 GW of generation and over 5,000 buses ranging from 69 to 765 kV, simulated in a TS-type RT simulator (RTS); (ii) realistic distribution feeder models with over 9,000 nodes each, simulated in a DSS, (iii) DERs such as photovoltaic (PV), energy storage (ES), water heaters, and HVAC, simulated in stand-alone DER simulator tools, and (iv) a DERMS responsible for issuing commands to the DER simulators to request the devices to provide grid services. 

Note OpenDERMS, displayed in Fig. \ref{fig:cosim_struct}, is a web application developed by the Electric Power Research Institute (EPRI). It can aggregate, control, and manage many DERs to provide grid support services. It supports various communication protocols such as the CTA-2045 standard with Energy Star specifications, DNP3, and SunSpec Modbus for large scale DER interoperability and evaluating different control strategies. In addition, the tool includes grid support function templates such as connect/disconnect, volt-var support, frequency-watt support, etc. It can be flexibly customized for testing several control functions and observing their impacts when co-simulated with DSS. For our testing methodology, the OpenDERMS tool was customized to receive time synchronization data via MQTT protocol and issue the control setpoints to the interconnected DERs via DNP3.

\begin{figure}[htb]
    \centering
    \includegraphics[width=0.5\textwidth]{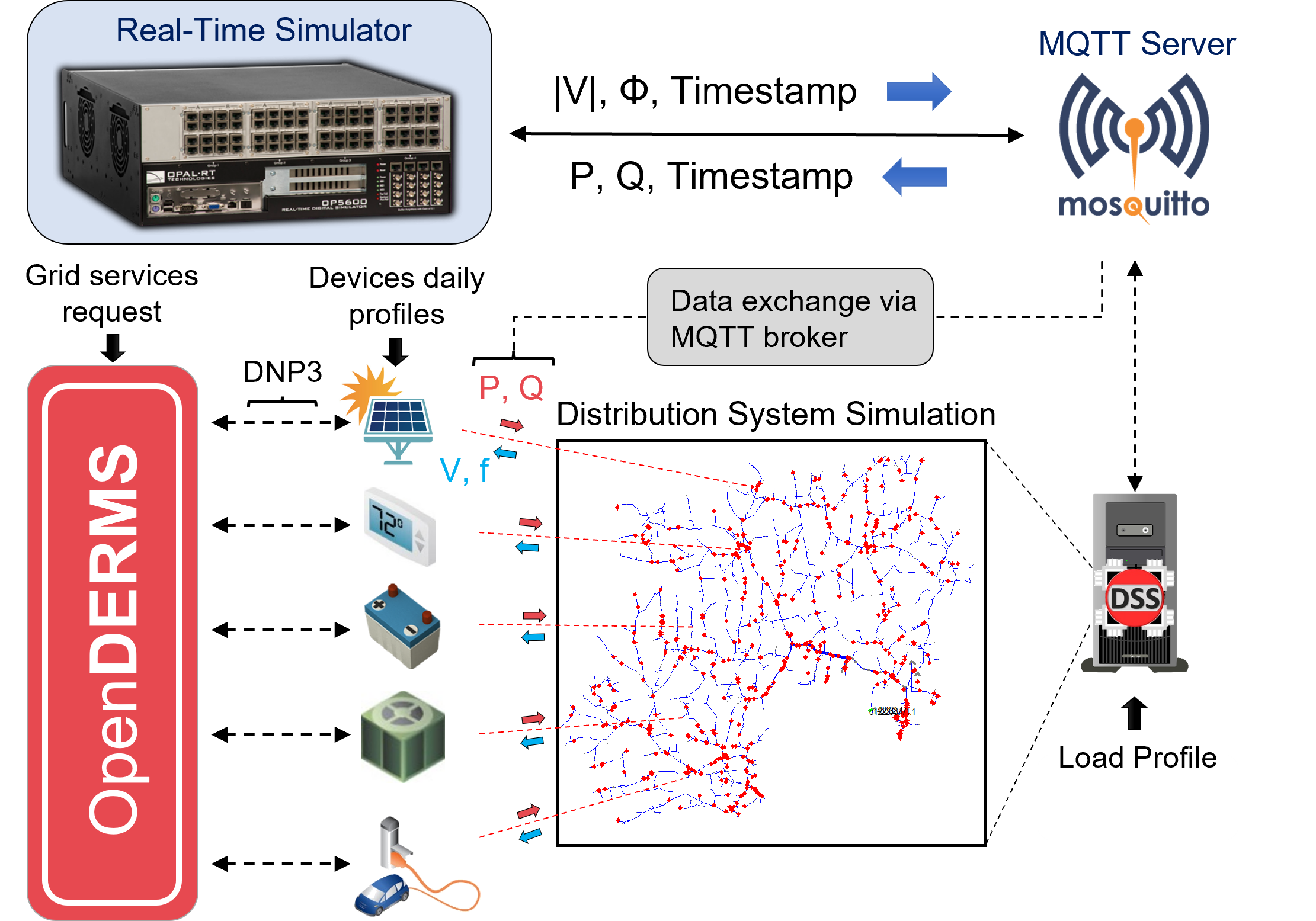}
    \caption{The power system variables exchanged between RTS, DSS, and OpenDERMS in one instance of the co-simulation loop}
    \label{fig:mqtt_connection}
\end{figure}

The communication architecture of the proposed RT T\&D co-simulation framework is based on MQTT communication protocol \cite{standard2014mqtt}, which is ideal for internet of things applications \cite{atmoko2017iot}. Figure \ref{fig:mqtt_connection} demonstrates how the data exchange between the RTS, the DSS, and the OpenDERMS platform is performed via the MQTT broker, which is an intermediary agent responsible for receiving messages published by clients, filtering messages by topic, and distributing them to subscribers. Note OpenDERMS is responsible for adjusting devices power setpoints based on grid service commands requested by grid operators. A more in-detail explanation of how the data exchange and time synchronization has been implemented in the proposed testbed is provided in \cite{paduani2022real}. 

\section{Simulation Results}

In this work, two main studies are presented: (A) an analysis of the DERs' impact on the system wide frequency response in cases with high DER penetration, and (B) an analysis of the capability of BTM DERs in tracking an AGC regulation signal. In both studies, 12 synchronous generators from the NYS BPS (equivalent to 2.878 GW) have been displaced by DERs to represent a future scenario of increased DER participation. For simplicity, the DERs consists only of PV generation.  

\subsection{Impact of DERs on System Wide Frequency Response}

\begin{figure}[htb]
    \centering
    \includegraphics[width=0.5\textwidth]{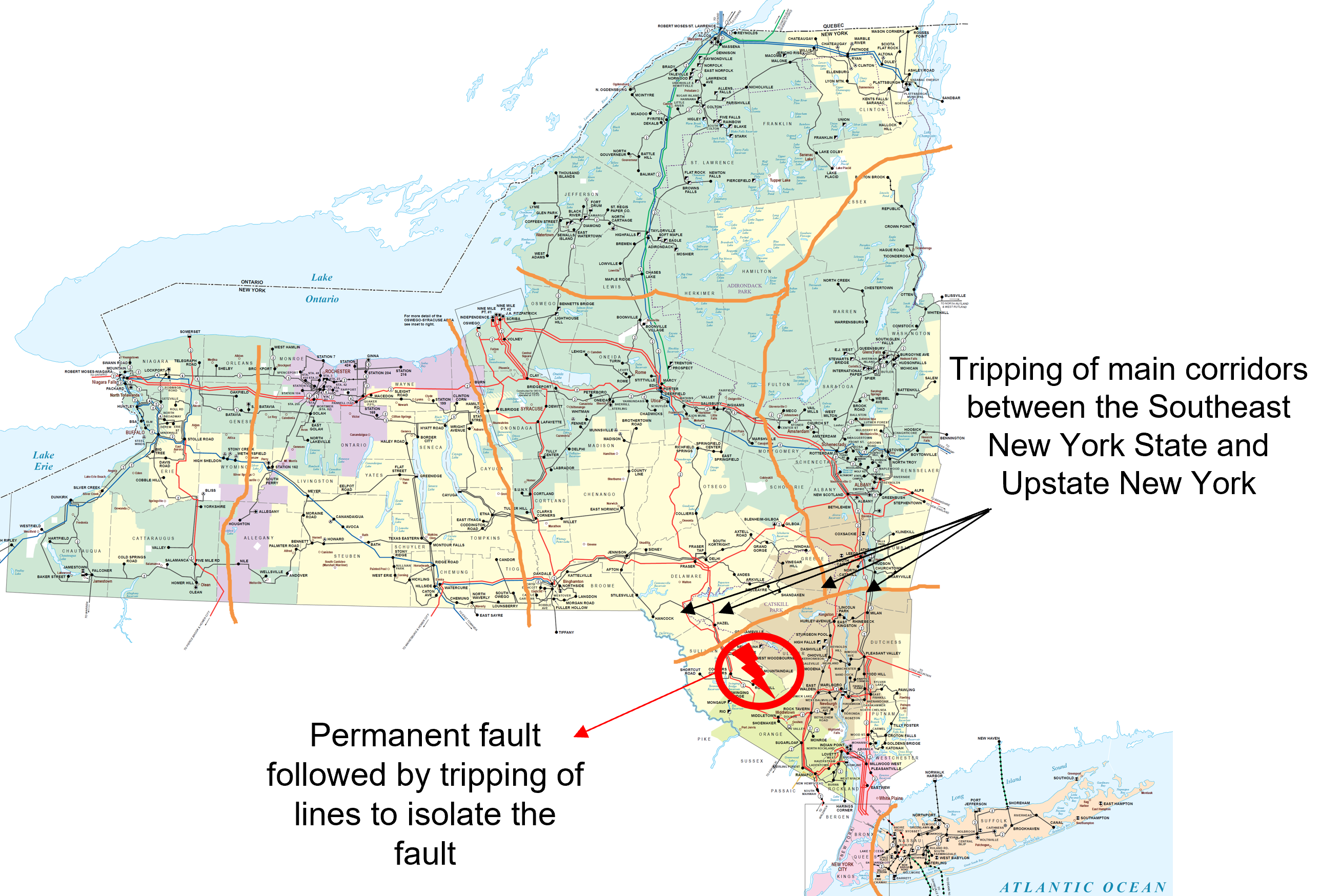}
    \caption{Three-phase to ground fault scenario studied.}
    \label{fig:trip_location}
\end{figure}

In this case, the RT T\&D co-simulation platform is composed by the 5,000 buses BPS and two 9,500-node distribution feeders with 500 DERs connected to each. A permanent three-phase to ground fault is applied to the transmission network, followed by the outage of subtransmission lines leading to the isolation of the fault as well as a load shedding of 400 MW just below the Central NYS. The event causes the tripping of the main corridors between the Southeast NYS and Upstate NY. Figure \ref{fig:trip_location} highlights the location of the fault. To highlight the impact of DERs on the system wide frequency response, the following use cases are subjected to the described scenario.

\begin{itemize}
    \item \textbf{Base Case:} represents a traditional power system. In this case, the synchronous generators are not substituted by DERs.
    \item \textbf{Case 1:} represents a power system with DERs not providing grid services. In this case, 2.878 GW of synchronous generation capacity from the NYS is replaced by DERs that operate in grid-following mode without frequency-watt droop response.
    \item \textbf{Case 2:} represents the future power system with DERs providing grid services required by standards \cite{ieee1547}. In this case, the same amount of synchronous generation from Case 1 is retired. However, the DERs are equipped with frequency-watt droop capabilities to help the grid during frequency events.
\end{itemize}

\begin{figure}[htb]
    \centering
    \includegraphics[width=0.5\textwidth]{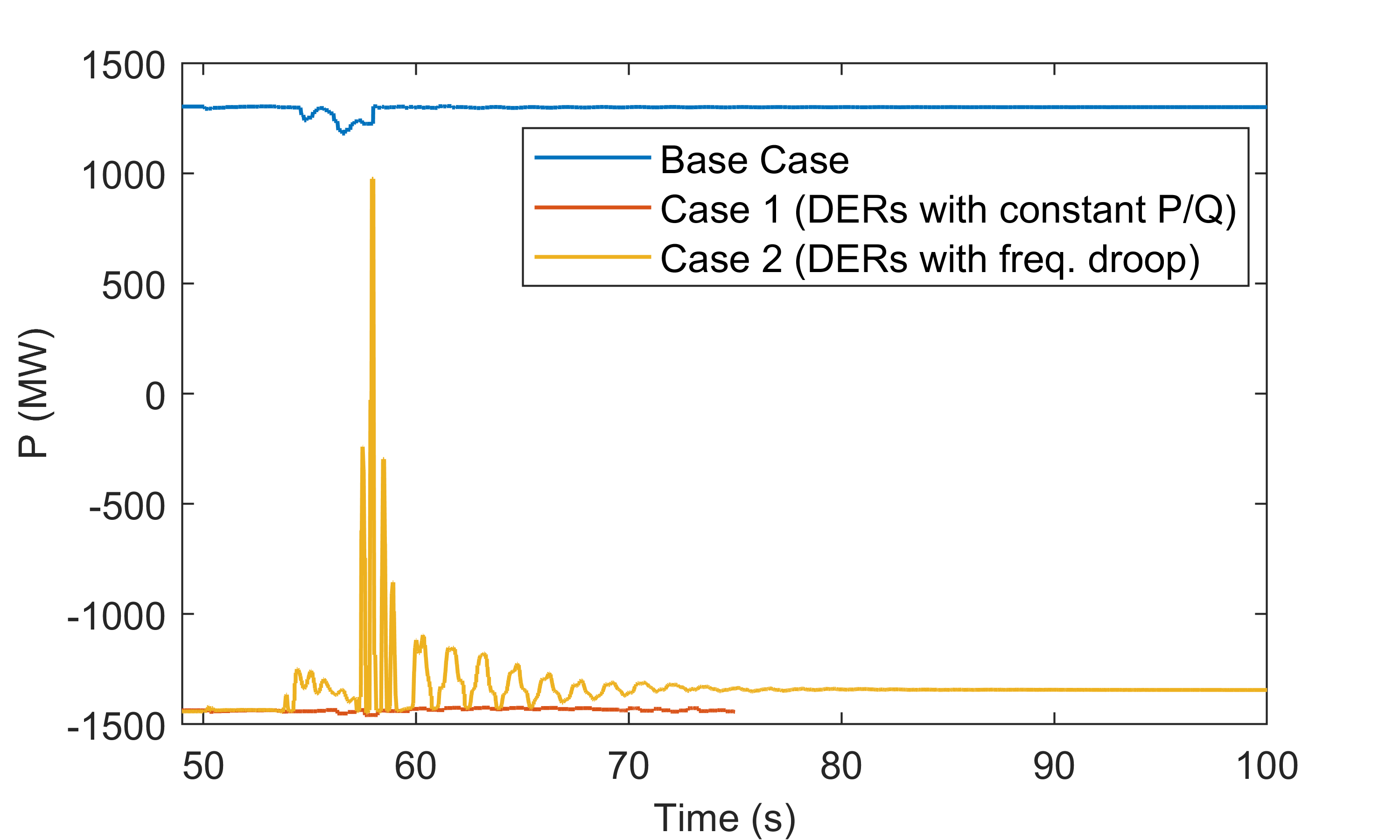}
    \caption{Total active power contribution from the distribution feeders hosting DERs.}
    \label{fig:active_power}
\end{figure}

\begin{figure}[htb]
    \centering
    \includegraphics[width=0.5\textwidth]{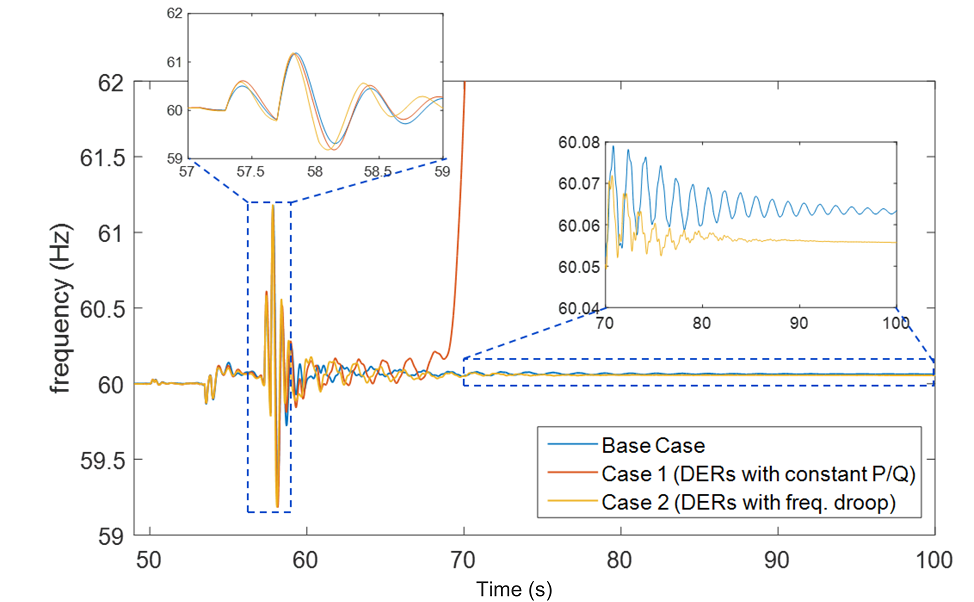}
    \caption{System frequency in response to the sudden load drop caused by the contingency.}
    \label{fig:frequency_response}
\end{figure}

Figure \ref{fig:active_power} displays the aggregated total active power consumption from the distribution feeders in each use case during the contingency scenario. Note cases 1 and 2 present negative power demand due to the DER power injection through the feeders. As seen in the figure, the power consumption is approximately constant in the Base Case and Case 1; however, it presents large oscillations in Case 2. This is because of the frequency droop controllers responsible for adjusting the output of the DERs in response to system frequency variations. Results show the DERs can considerably assist the grid during the fault event in Case 2. Figure \ref{fig:frequency_response} displays the system frequency for each use case.

Although almost 3 GW of synchronous generation was replaced by DERs in Cases 1 and 2, the total system inertia was not significantly impacted. This can be observed by comparing the frequency response during the fault recovery between the use cases. During the first minute of simulation, all three cases present similar frequency oscillations, ranging between approximately 59.2 to 61.25 Hz. This is because the transmission system in study corresponds to a section of the whole Eastern Interconnection grid, with very high inertia. 

As shown in Fig. \ref{fig:frequency_response}, the system is able to maintain stability after the fault in the Base Case, in which no synchronous machine was retired. Nevertheless, the lack of any frequency response from the DERs in Case 1 caused the system to oscillate and lose stability around 70 seconds of simulation. Noticeably, simulation results between 70 to 100 seconds of simulation demonstrate how DERs equipped with frequency droop controllers (Case 2) can not only damp the post-event oscillations but also help the frequency to settle at a steady-state closer to its nominal value when compared to the Base Case. This example highlights the importance of having DERs following grid standards such as IEEE 1547-2018 to ensure grid stability during contingency events. It is worth mentioning that the grid frequency in display has been measured at a bus in the central NYS region.

\subsection{AGC Signal Tracking with DERs}

In this case, the RT T\&D co-simulation platform is composed by the 5,000 buses BPS and 10 distribution systems connected to different buses of the BPS, in which each distribution system corresponds to 92 IEEE 123 node test feeders connected in parallel. The same amount of synchronous machine generation from the previous study is retired, such that the total DER power injection (represented as PV systems) corresponds to 2.878 GW when the PVs operate at rated power. The reason behind having specifically 92 feeders in parallel is to have the overall distribution system output matching the power injection from the retired synchronous machines. Since there is a limit for the PV hosting capacity before reaching overvoltage issues, the number of feeders in parallel had to be increased instead of adding more DERs on the same feeder. 

The PV systems are set to follow AGC regulation signal. Moreover, to provide both upward and downward AGC tracking capability, it is assumed the PV systems operate with headroom, which can be achieved via maximum power point estimation algorithms \cite{unified2021paduani}. Note the DER simulators do not consider PV power setpoint tracking algorithms, and the results correspond to moments when the PV available power was above or equal the AGC requested setpoint. Therefore, in this case, the DERs can track the AGC power setpoints without errors. The following use cases are established to compare the AGC signal tracking capability between utility-scale and small BTM DERs.

\begin{itemize}
    \item \textbf{Case 1:} represents today's power system, in which large utility-scale DERs participate in the transmission market, while small-scale DERs do not. In this case, a three-phase 6.85 MW utility-scale PV plant is sited at the feeder head of each IEEE 123 node test feeder, corresponding to the low-voltage side of the substation transformer.
    \item \textbf{Case 2:} demonstrates AGC signal tracking by small-scale DERs managed by aggregators. In this case, the utility-scale PV plant at the feeder head is substituted by 100 BTM, single-phase PV systems distributed through each IEEE 123 node test feeder, each unit with 68.5 kW rated power. 
\end{itemize}

\begin{figure}[htb]
    \centering
    \includegraphics[width=0.5\textwidth]{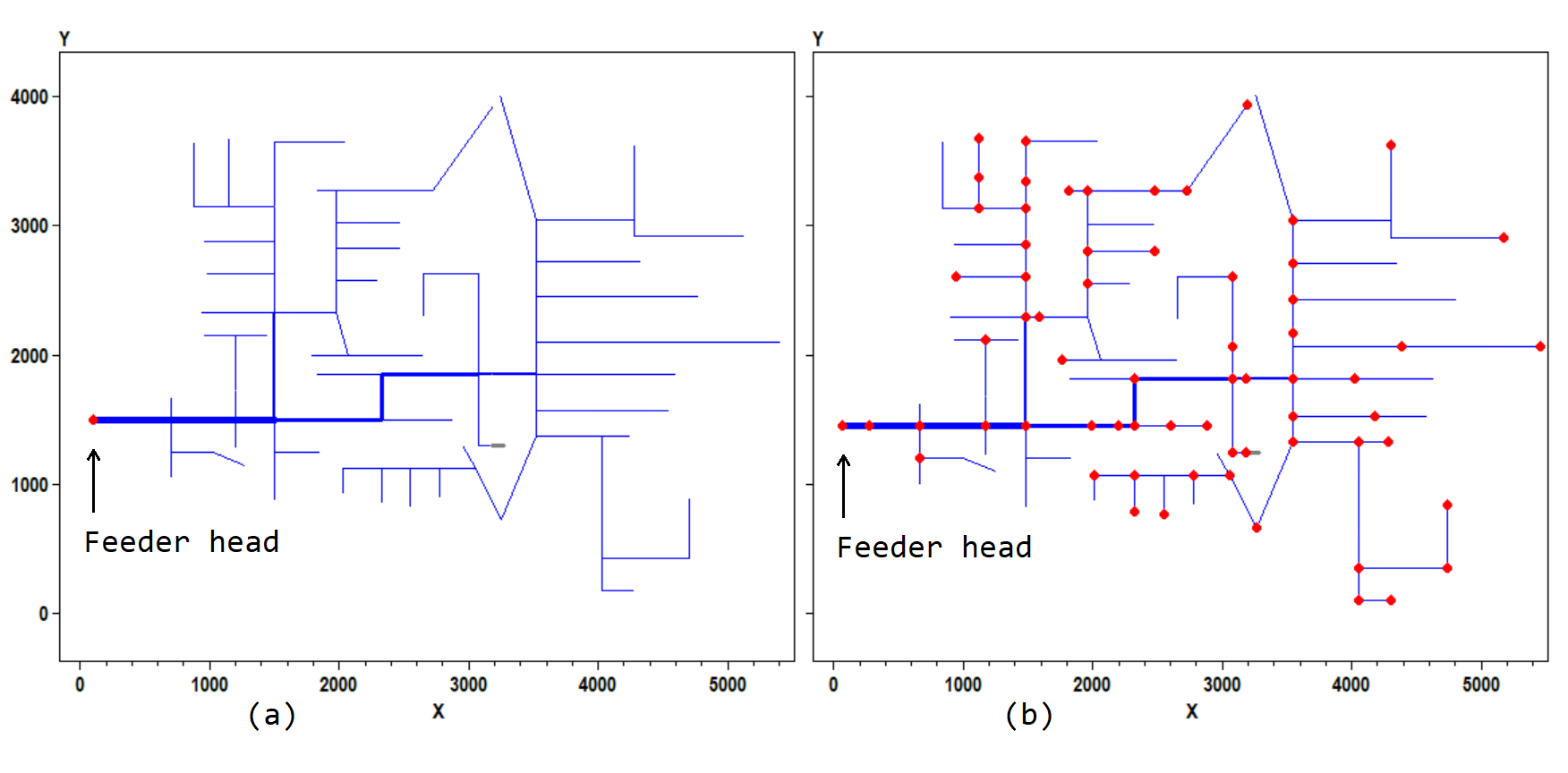}
    \caption{DER allocation (indicated by red dots) inside the IEEE 123 node test feeder: (a) Case 1 with a utility-scale PV plant at the feeder head, (b) Case 2 with 100 BTM PVs distributed through the feeder.}
    \label{fig:feederfig2}
\end{figure}

Figure \ref{fig:feederfig2} shows the allocation of DERs inside the distribution feeder in Cases 1 and 2. Note each red marker indicating a DER unit in Fig. \ref{fig:feederfig2}(b) may correspond to PVs connected to more than one phase. Both cases contain the same PV capacity.

\begin{figure}[htb]
    \centering
    \includegraphics[width=0.5\textwidth]{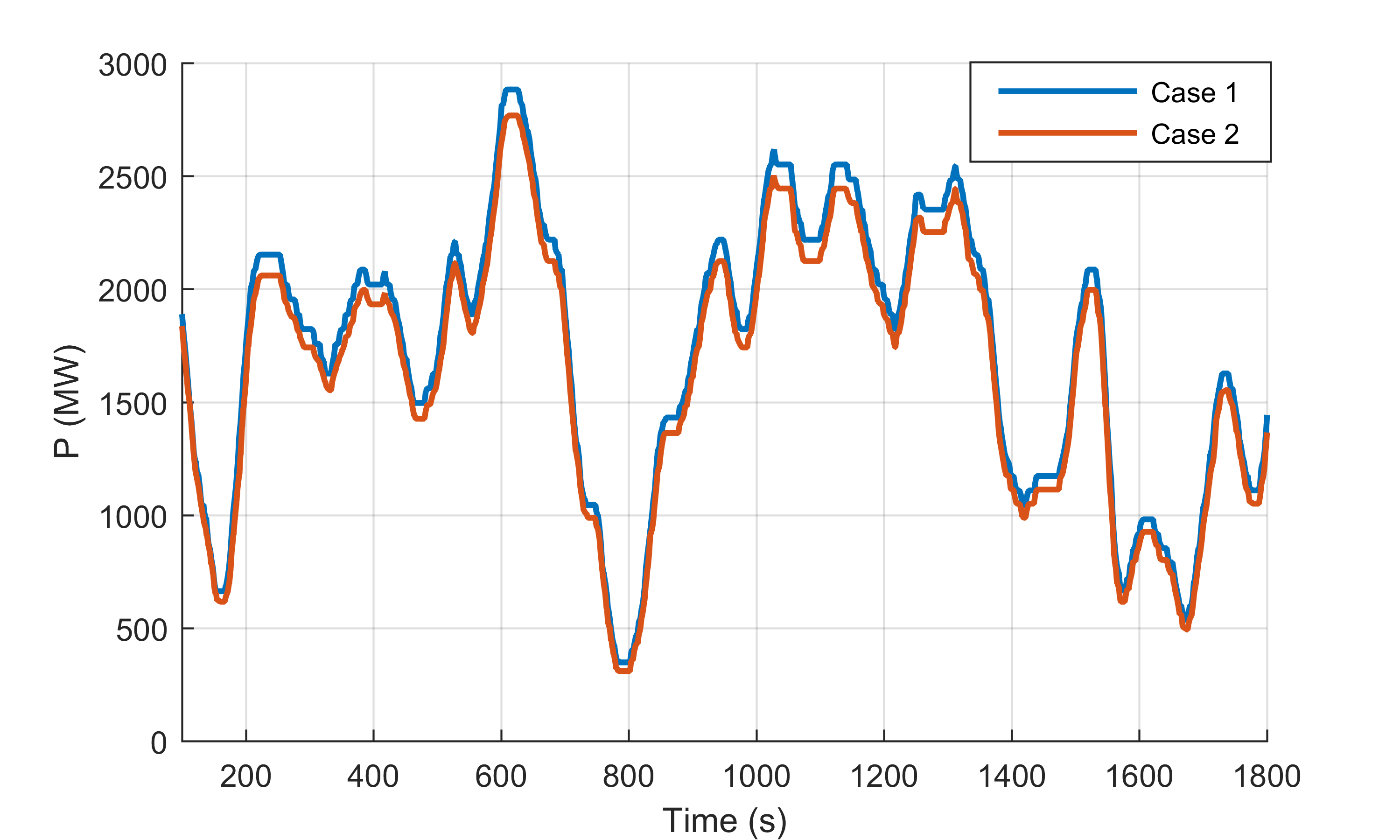}
    \caption{Comparison of the overall power injection into the transmission system between Cases 1 and 2.}
    \label{fig:AGC_power}
\end{figure}

A comparison of the DERs output power between Cases 1 and 2 is displayed in Fig. \ref{fig:AGC_power}. Note the PV systems receive the power setpoint from the DERMS and inject their power into their corresponding bus in the IEEE 123 node test feeder (Fig. \ref{fig:feederfig2}). But the power displayed in the figure represents the sum of powers flowing from all feeders being simulated, and it is measured at the feeder head at the high voltage side of the substation transformer. 

\begin{figure}[htb]
    \centering
    \includegraphics[width=0.5\textwidth]{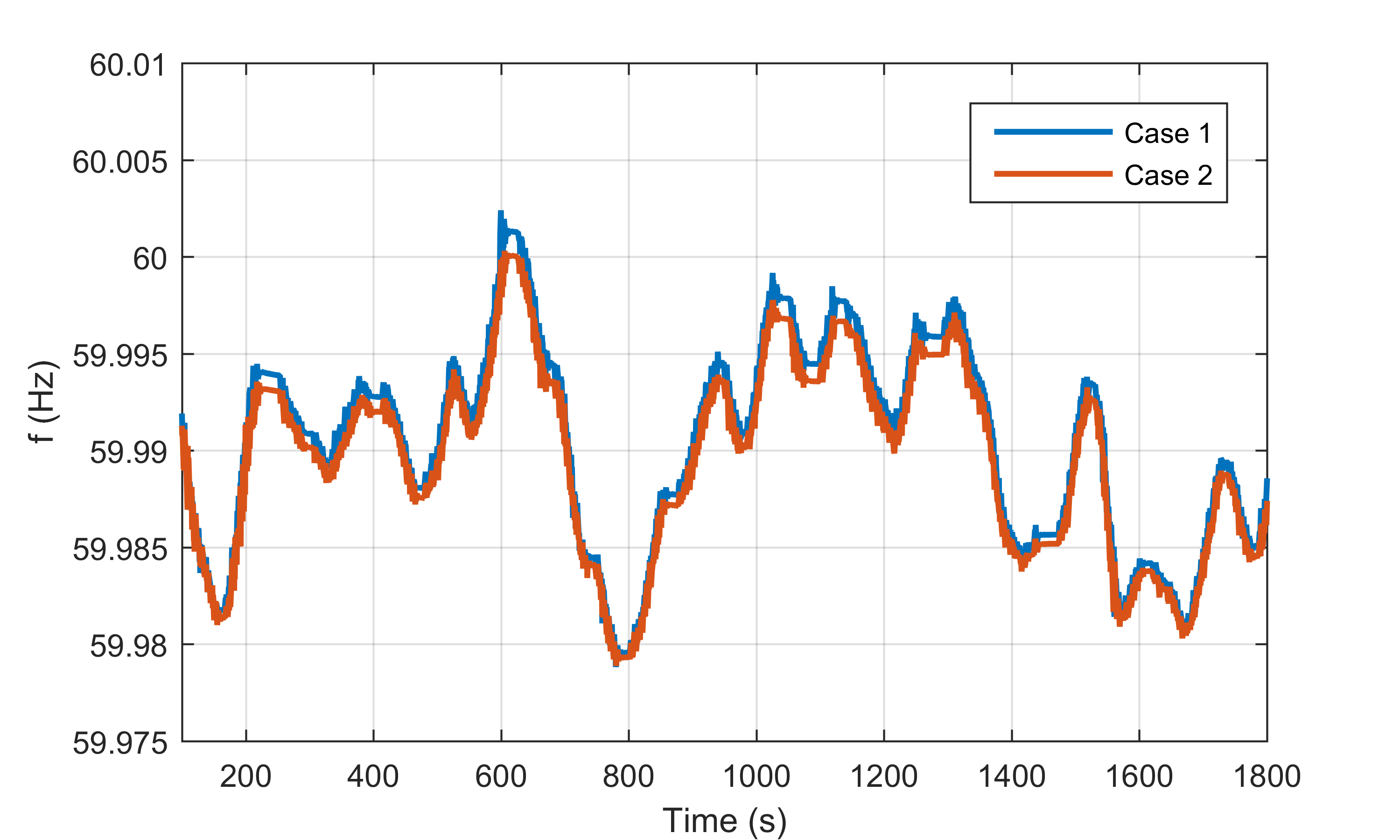}
    \caption{Comparison of the BPS frequency between Cases 1 and 2.}
    \label{fig:freqreg}
\end{figure}

\begin{figure}[htb]
    \centering
    \includegraphics[width=0.5\textwidth]{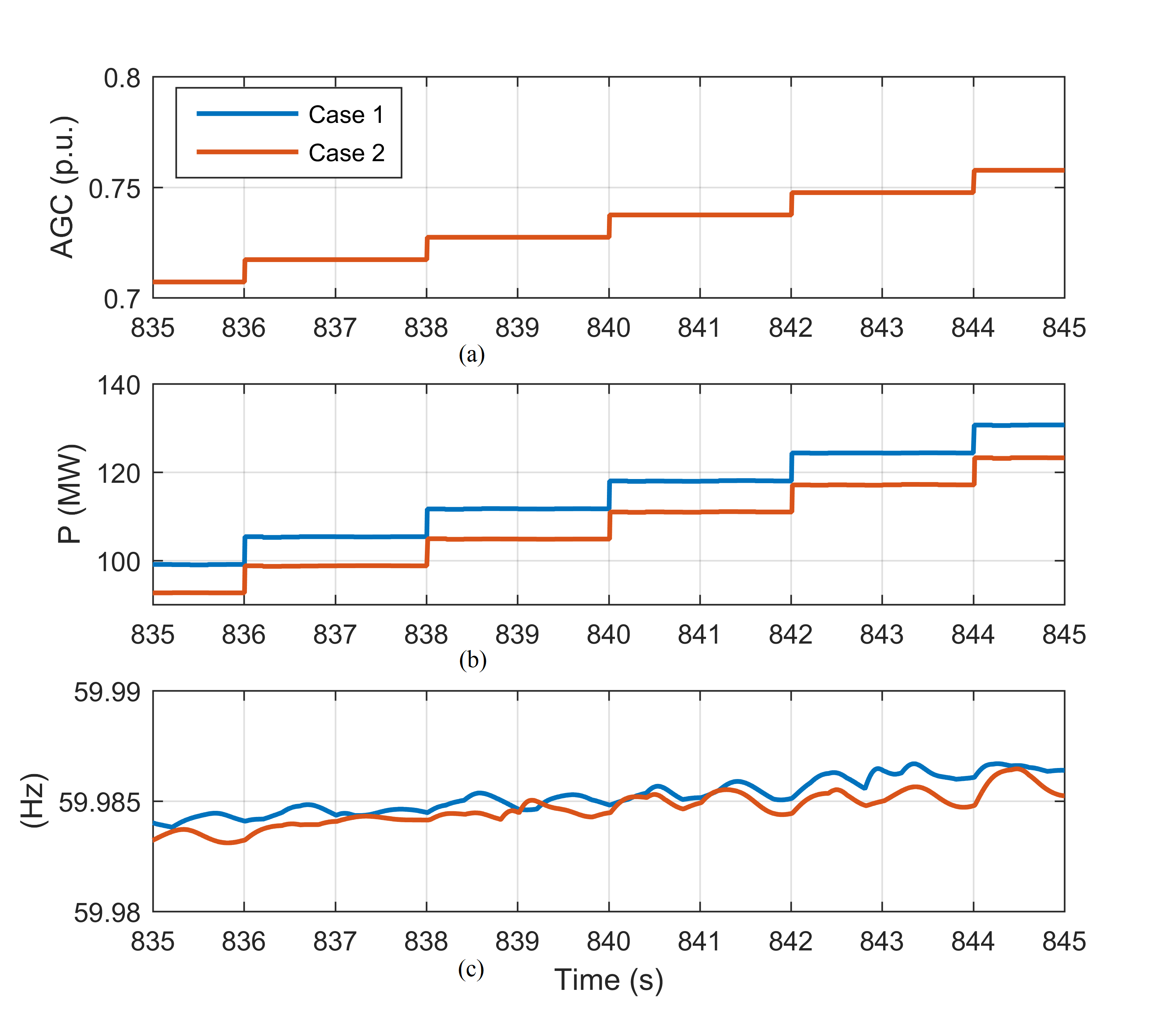}
    \caption{Snapshot of the AGC signal tracking for each case: (a) AGC signal sent by the DERMS, (b) power injection at one bus of the transmission system, and (c) measured BPS frequency.}
    \label{fig:AGC_zoom}
\end{figure}

Even though both cases have the same allocated PV capacity, Case 2 presents a smaller power output when compared to Case 1. That difference is caused by losses through the feeder, since the utility-scale PVs in Case 1 are allocated at the feeder head, and have a shorter path towards the transmission system. Figure \ref{fig:freqreg} displays the BPS frequency for both cases, whereas Fig. \ref{fig:AGC_zoom} displays a snapshot of the AGC signal, the corresponding output power flowing to one of the transmission system (note there are 10 different buses, each connected to an identical distribution system), and the BPS frequencies for each case. Despite the small output power offset between the cases, results demonstrate that when DERMS are utilized to control BTM DERs via Aggregators, small DERs can also be utilized to provide grid services, reaching a similar level of frequency regulation capability when compared to utility-scale systems. Note the BPS frequency from the TS-type RTS is obtained via generator slip measurements.

\begin{figure}[htb]
    \centering
    \includegraphics[width=0.5\textwidth]{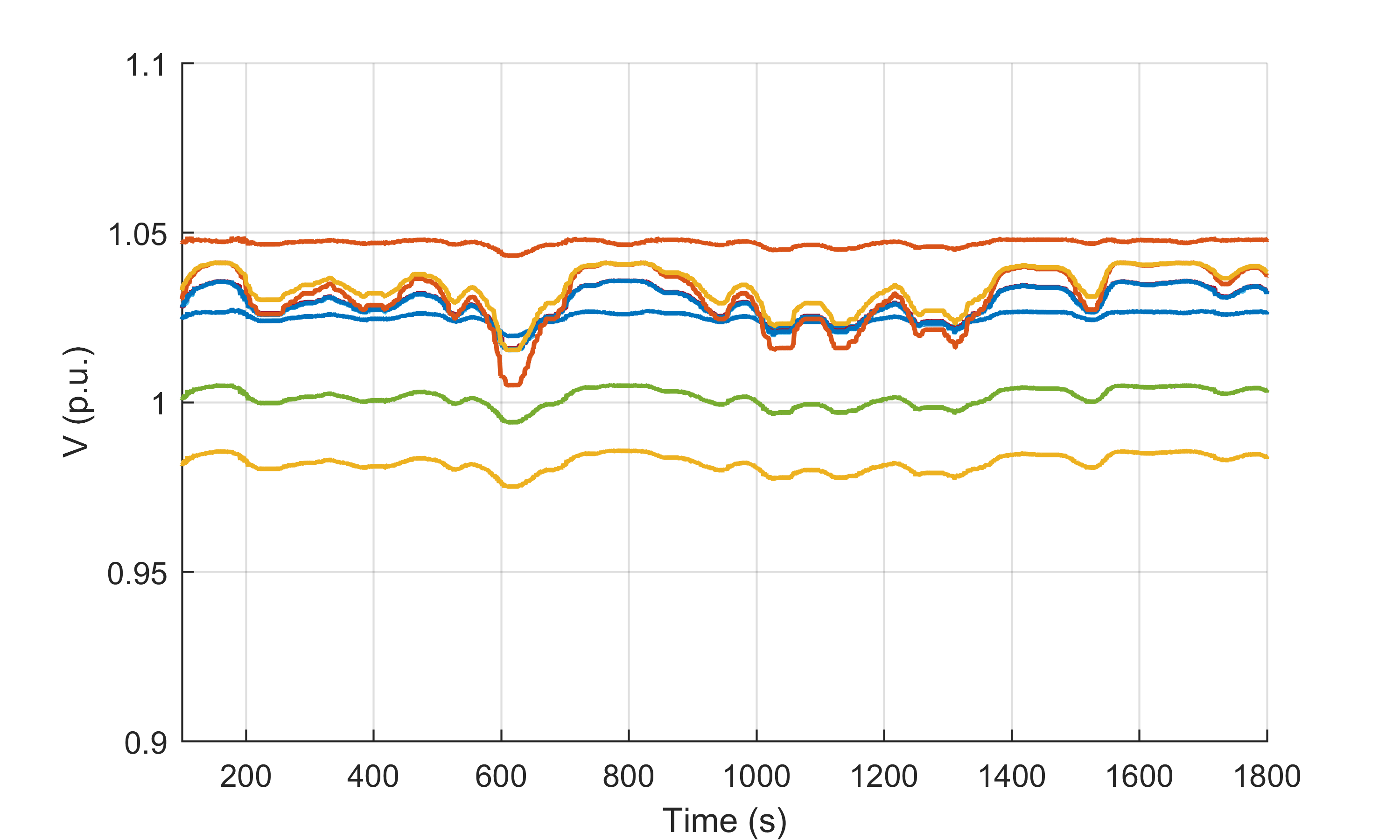}
    \caption{Voltages measured at the higher side of the substation transformer for each of the 10 transmission buses connected to distribution systems in Case 2.}
    \label{fig:voltages_case2}
\end{figure}

The oscillations in DER power injection also impact the system voltages. Figure \ref{fig:voltages_case2} displays the voltages measured at the higher side of the substation transformer for each of the 10 transmission buses connected to distribution systems. As shown in the figure, the system voltages are maintained within acceptable ranges while the DERs follow the AGC tracking. However, changes in PV injection will also affect the operation of the distribution feeder, adding unwanted stress to voltage regulation devices that may reduce their lifespan. Future studies with the co-simulation testbed will be carried out to analyze the impact of DER power variations on the distribution system operation.

\section{Conclusion}

This paper investigates the capability of BTM DERs in providing grid services in large-scale T\&D systems. By combining a TS-type RTS, a DSS, and DER simulators controlled by DERMS via MQTT communication protocol, the proposed RT T\&D co-simulation testbed presents itself as an excellent tool for simulating grids under high penetration of DERs and testing DERMS algorithms. In the first study, it is found that when large synchronous generators are substituted by DERs without GSFs, the BPS can reach instability if subjected to severe contingencies. Yet, if the DERs are equipped with GSFs, they can assist damping frequency oscillations, helping to maintain system stability. Moreover, the second study analyzes the capability of BTM DERs in participating in the transmission market by tracking AGC regulation signals. Results confirmed their potential to provide a similar level of grid support expected from utility-scale DERs. Future work on this topic will evaluate the impact of DERs providing transmission services on the distribution system operation.


\bibliographystyle{IEEEtran}
\bibliography{mybibtex}

\begin{thebibliography}{10}
\providecommand{\url}[1]{#1}
\csname url@samestyle\endcsname
\providecommand{\newblock}{\relax}
\providecommand{\bibinfo}[2]{#2}
\providecommand{\BIBentrySTDinterwordspacing}{\spaceskip=0pt\relax}
\providecommand{\BIBentryALTinterwordstretchfactor}{4}
\providecommand{\BIBentryALTinterwordspacing}{\spaceskip=\fontdimen2\font plus
\BIBentryALTinterwordstretchfactor\fontdimen3\font minus
  \fontdimen4\font\relax}
\providecommand{\BIBforeignlanguage}[2]{{%
\expandafter\ifx\csname l@#1\endcsname\relax
\typeout{** WARNING: IEEEtran.bst: No hyphenation pattern has been}%
\typeout{** loaded for the language `#1'. Using the pattern for}%
\typeout{** the default language instead.}%
\else
\language=\csname l@#1\endcsname
\fi
#2}}
\providecommand{\BIBdecl}{\relax}
\BIBdecl

\bibitem{cano2020ferc}
C.~Cano, ``{FERC Order No. 2222: A New Day for Distributed Energy Resources},''
  2020.

\bibitem{EPRI2022FERC}
``{DER} aggregation participation in electricity markets: {EPRI} collaborative
  forum final report and {FERC} order 2222 roadmap,'' Electric Power Research
  Institute ({EPRI}), Palo Alto, CA (United States), Tech. Rep., 2022.

\bibitem{aminul2021grid}
N.~Singhal, M.~Heidarifar, T.~Hubert, E.~Ela, A.~Huque, T.~Abate, and P.~Shoop,
  ``Grid services in the distribution and bulk power systems,'' Electric Power
  Research Institute.(EPRI), Palo Alto, CA (United States), Tech. Rep., 2021.

\bibitem{ieee1547}
``{IEEE Standard 1547-2018},'' \emph{Standard for interconnection and
  interoperability of distributed energy resources with associated electric
  power systems interfaces}, 2018.

\bibitem{calrule21}
{California Public Utilities Commission}, ``{Electric Rule No. 21 Generating
  facility interconnections},'' 2016.

\bibitem{aminul2021distributed}
A.~Garg, T.~Hubert, A.~Huque, and A.~Renjit, ``Distributed energy resource
  magagement system ({DERMS}) control architecture for grid services in the
  distribution and bulk power systems.'' Electric Power Research
  Institute.(EPRI), Palo Alto, CA (United States), Tech. Rep., 2021.

\bibitem{matevosyan2021future}
J.~Matevosyan, J.~MacDowell, N.~Miller, B.~Badrzadeh, D.~Ramasubramanian,
  A.~Isaacs, R.~Quint, E.~Quitmann, R.~Pfeiffer, H.~Urdal \emph{et~al.}, ``A
  future with inverter-based resources: Finding strength from traditional
  weakness,'' \emph{IEEE Power and Energy Magazine}, vol.~19, no.~6, pp.
  18--28, 2021.

\bibitem{jain2021integrated}
H.~Jain, B.~A. Bhatti, T.~Wu, B.~Mather, and R.~Broadwater, ``Integrated
  transmission-and-distribution system modeling of power systems:
  State-of-the-art and future research directions,'' \emph{Energies}, vol.~14,
  no.~1, p.~12, 2021.

\bibitem{palmintier2017design}
B.~Palmintier, D.~Krishnamurthy, P.~Top, S.~Smith, J.~Daily, and J.~Fuller,
  ``Design of the {HELICS} high-performance
  transmission-distribution-communication-market co-simulation framework,'' in
  \emph{2017 Workshop on Modeling and Simulation of Cyber-Physical Energy
  Systems (MSCPES)}.\hskip 1em plus 0.5em minus 0.4em\relax IEEE, 2017, pp.
  1--6.

\bibitem{poudel2022modeling}
S.~Poudel, S.~J. Keene, R.~L. Kini, S.~Hanif, R.~B. Bass, and J.~T. Kolln,
  ``Modeling environment for testing a distributed energy resource management
  system (derms) using gridapps-d platform,'' \emph{IEEE Access}, vol.~10, pp.
  77\,383--77\,395, 2022.

\bibitem{paduani2022real}
V.~Paduani, R.~Kadavil, H.~Hooshyar, A.~Haddadi, A.~Jakaria, and A.~Huque,
  ``Real-time {T\&D} co-simulation for testing grid impact of high {DER}
  participation,'' in \emph{2022 IEEE Power \& Energy Society Grid Edge
  (PESGE)}.\hskip 1em plus 0.5em minus 0.4em\relax IEEE, 2022.

\bibitem{standard2014mqtt}
\BIBentryALTinterwordspacing
``{MQTT} version 3.1.1.'' [Online]. Available:
  \url{http://docs.oasis-open.org/mqtt/mqtt/v3}
\BIBentrySTDinterwordspacing

\bibitem{atmoko2017iot}
R.~Atmoko, R.~Riantini, and M.~Hasin, ``{IoT} real time data acquisition using
  {MQTT} protocol,'' in \emph{Journal of Physics: Conference Series}, vol. 853,
  no.~1.\hskip 1em plus 0.5em minus 0.4em\relax IOP Publishing, 2017, p.
  012003.

\bibitem{unified2021paduani}
V.~Paduani, H.~Yu, B.~Xu, and N.~Lu, ``A unified power-setpoint tracking
  algorithm for utility-scale {PV} systems with power reserves and fast
  frequency response capabilities,'' \emph{IEEE Transactions on Sustainable
  Energy}, vol.~13, no.~1, pp. 479--490, 2021.

\end{thebibliography}

\end{document}